\documentclass[journal]{IEEEtran}
\usepackage{cite}
\usepackage{amsmath,amssymb,amsfonts,esint}
\usepackage{algorithmic}
\usepackage{graphicx}
\usepackage{textcomp}

\newcommand{\E}{\ensuremath{\mathrm{e}}}        
\newcommand{\degree}{\ensuremath{^\circ}}

\providecommand{\D}{\,\mathrm{d}}               
\providecommand{\M}[1]{\mathbf{#1}}             
\providecommand{\T}[1]{\mathrm{#1}}             

\providecommand{\herm}{\mathrm{H}}

\def\BibTeX{{\rm B\kern-.05em{\sc i\kern-.025em b}\kern-.08em
    T\kern-.1667em\lower.7ex\hbox{E}\kern-.125emX}}

\begin{document}
%

\title{On the End-Fire (Super) Directivity of an Array of Two Elementary Dipoles}
\author{Pavel Hazdra, \IEEEmembership{Member, IEEE}, Jan Kracek, \IEEEmembership{Member, IEEE}, and Tomas Lonsky
\thanks{The authors are with the Czech Technical University in Prague, Faculty of Electrical Engineering, Department of Electromagnetic Field, Czech Republic (e-mail: hazdrap@fel.cvut.cz).}
}
\maketitle

\markboth{Journal of \LaTeX\ Class Files,~Vol.~14, No.~8, August~2015}%
{Shell \MakeLowercase{\textit{et al.}}: Bare Demo of IEEEtran.cls for IEEE Journals}

\maketitle

\begin{abstract}
The concept of source currents of a radiating source can be employed to express directivity in some particular cases analytically. For an antenna array, this concept can be combined with concepts of mutual radiation intensity and the mutual power of array elements. Using these approaches, we treat the evaluation of the directivity for an elementary dipole, and its end-fire array, which is of special attention due to superdirective properties. A closed form expression for the directivity of this array with \mbox{out-of-phase} excitation is derived. It is observed that end-fire directivity can be further enhanced by optimizing the excitation currents of the array. Their optimal relative phase and corresponding increased directivity are also found analytically. The results are validated by a full-wave simulator.
\end{abstract}

\begin{IEEEkeywords}
directivity, superdirectivity, antenna array, elementary dipole, optimal excitation
\end{IEEEkeywords}

%
\IEEEpeerreviewmaketitle

\section{Introduction}
\label{sec:introduction}
\IEEEPARstart{I}{t}
is well known that an end-fire antenna array of closely-spaced elements is able to show a significant increase in directivity (termed superdirectivity) compared to a sole element \cite{Hansen_Woody_Directional_Antenna_Design},\cite{Bloch_superdirectivity_1953}. Uzkov derived the end-fire directivity limit for the case of $N$ isotropic radiators, when directivity approaches $N^2$ as the distance between them reaches zero \cite{Uzkov}. For more general multipole radiators, Harrington obtained the limit to be $N^2+2N$ \cite{Harrington_OnTheGainAndBWofDirectAntennas}.
Recently, the design of closely spaced array elements (when the distance is less than  $\lambda/4$, where $\lambda$ is wavelength) attracted both theoretical and practical interest \cite{YaghSmallSupergain}, \cite{Haskou},\cite{Altshuler},\cite{Noguchi3el}, \cite{Best_ImpedanceMatched}, \cite{Clemente_SuperD}. We should also mention the first realization of such an array, the Kraus’s W8JK antenna\cite{Kraus_Antennas}.

In this paper, we firstly derive a generalized directivity of a radiating source, the antenna array,  based on its current distributions and excitation currents. This results in a framework of self- and mutual intensities and self- and mutual radiated powers of array elements which is similar to the approach developed by Hansen who used mutual radiation resistances in his derivations array directivies \cite{RudgeHandbookVol2}.

In the case of elementary dipoles, which are considered here as the array elements, the integrals contained in the relation for the directivity are easy to work out in the closed form. A simple formula for the directivity of the array of two out-of-phase excited dipoles spaced less than $\lambda/2$ is obtained. Furthermore, the quadratic form of the excitation currents involved allows, by means of the generalized eigenvalue problem, the optimum to be found, thus producing a maximal directivity of this configuration. The optimum is also derived in the closed form by following the approach of Uzsoky and Solymar \cite{UzsokySolymar_TheoryOfSuperDirectiveLinearArrays}. In this manner, the ``superdirective factor" of $21/15$, accounting for the increased directivity between the optimal and out-of-phase excitation, is found. A similar factor of $4/3$ is discovered for an array of two isotropic radiators.

The results are verified by a full-wave simulator CST \cite{cst} and show good agreement.

\section{Directivity in Terms  of  Source Currents}
A concept of the generalized directivity of a radiating source, the antenna array, based on its current distributions and excitation currents, is reviewed in this section.

A directivity of a radiating source in an angular direction $(\theta,\phi)$ in the spherical coordinates is defined as \cite{Orfanidis_ElectromagneticWaves}
\begin{equation}\label{eq:Dir}
	D(\theta,\phi)=\frac{U(\theta,\phi)}{U_0}=4\pi\frac{U(\theta,\phi)}{P_{\mathrm r}}
\end{equation}
where $U$ is radiation intensity in the direction $(\theta,\phi)$, $U_0=P_{\mathrm r}/4\pi$ is average radiation intensity and $P_{\mathrm r}$ is radiated power. Intensity $U$ is related to a far electric field $\M{E}_{\mathrm {far}}$ of the source as
\begin{equation}\label{eq:U}
	U(\theta,\phi)=r^2S_{\mathrm r}=r^2\frac{|\M{E}_{\mathrm {far}}(r,\theta,\phi)|^2}{2Z_0}
\end{equation}
where $r$ is distance from the origin of the coordinates, $S_{\mathrm r}$ is radial power density and $Z_0=120\pi$ is an impedance of free space. The far electric field $\M{E}_{\mathrm {far}}$ may be expressed as
\begin{equation}\label{eq:E}
	\M{E}_{\T {far}}(r,\theta,\phi)=\j\omega\M{r}_0\times(\M{r}_0\times\M{A}_{\T {far}}(r,\theta,\phi))
\end{equation}
where $\M{A}_{\mathrm {far}}$ is a vector potential of the source given by
\begin{equation}\label{eq:A}
	\M{A}_{\mathrm {far}}(r,\theta,\phi)=\frac{\mu_0}{4\pi}\frac{\E^{-\j kr}}{r}\int\limits_{V} \M{J}(\M{r}')\E^{\j k\Delta'}\D \M{r}'.
\end{equation}
In the above equation, the integration is performed over a (finite) volume $V$ of a current density $\M{J}$ of the source. Furthermore, $\mu_0$ is a permeability of vacuum and $\Delta'=\M{r}_0\cdot \M{r}'$ where a unit vector $\M{r}_0=[\sin(\theta) \cos(\phi), \sin(\theta) \sin(\phi), \cos(\theta)]$ determines the direction of radiation and the radius vector $\M{r}'=[x',y',z']$ describes the location of current $\M{J}$.

In the case of the source represented by an array of $N$ elements, current $\M{J}$ can be written as
\begin{equation}\label{eq:J}
	\M{J}(\M{r})=\sum_{n=1}^{N} \M{J}_n(\M{r})=\sum_{n=1}^{N} I_n\M{j}_n(\M{r})
\end{equation}
where $\M{J}_n$ is a current density existing in a volume $V_n$ of the $n$-th element and $\M{j}_n$ is a current density normalized to its excitation current $I_n$.
By inserting (\ref{eq:J}) through (\ref{eq:A}) and (\ref{eq:E}) into (\ref{eq:U}) and using $|\M{J}|^2=\M{J}\cdot \M{J}^*$, we arrive at the expression
\begin{equation}\label{eq:Usum}
U(\theta,\phi)=\sum_{m=1}^{N}\sum_{n=1}^{N} U_{mn}(\theta,\phi)
	\end{equation}
where
\begin{equation}\label{eq:Umn}
	\begin{split}
		U_{mn}(\theta,\phi)&=I_mI_n^*\frac{15k^2}{4\pi}\int\limits_{V_m}\int\limits_{V_n}\Lambda(\M{r},\M{r}')\E^{\j k(\Delta-\Delta')}\D \M{r}\D \M{r}'\\
						   &=I_mI_n^*u_{mn}(\theta,\phi)
	\end{split}
\end{equation}
is a mutual radiation intensity that accounts for the interaction of the $m$-th and $n$-th elements and $u_{mn}$ is its normalization to the currents $I_m$ and $I_n$. Furthermore, $k=2\pi/\lambda$ is a wavenumber,
\begin{equation}\label{eq:delta}
	\begin{split}
		\Delta-\Delta'=\M{r_0}\cdot(\M{r}-\M{r}')
		=&(x-x')\sin\theta\cos\phi\\
		&+(y-y')\sin\theta\sin\phi\\
		&+(z-z')\cos\theta,	
	\end{split}
\end{equation}
\begin{equation}\label{eq:Lambda}
	\Lambda(\M{r},\M{r}')=\M{j}_m(\M{r})\cdot\M{j}_n^*(\M{r}')-\M{r}_0 \cdot \M{j}_m(\M{r})  \M{r}_0 \cdot \M{j}_n^\ast(\M{r}'),
\end{equation}
see Appendix A\footnote{For the sake of simplicity, we do not treat intensity $U_{mn}$ for vertical and horizontal polarization separately in this paper.}.

The radiated power $P_{\mathrm{r}}$ required for calculating directivity $D$ (\ref{eq:Dir}) can be obtained through the EMF method \cite{JordanBalmainEM}, or by integrating intensity $U$ (\ref{eq:Usum}) over the complete solid angle
\begin{equation}\label{eq:Prad}
	P_{\mathrm{r}}=\int\limits_{0}^{2\pi}\int\limits_{0}^{\pi} U(\theta,\phi)\sin\theta\D\theta\D\phi
	=\sum_{m=1}^{N}\sum_{n=1}^{N} P_{mn}
\end{equation}
where
\begin{equation}\label{eq:Pmn}
	P_{mn}=I_mI_n^*\int\limits_{0}^{2\pi}\int\limits_{0}^{\pi} u_{mn}(\theta,\phi)\sin\theta\D\theta\D\phi=I_mI_n^*p_{mn}
\end{equation}
is a mutual power of the $m$-th and $n$-th elements and $p_{mn}$ is its normalization to the currents $I_m$ and $I_n$.

Consequently, directivity $D$ (\ref{eq:Dir}) takes a compact matrix form using (\ref{eq:Usum}), (\ref{eq:Umn}), (\ref{eq:Prad}), (\ref{eq:Pmn})  \cite{UzsokySolymar_TheoryOfSuperDirectiveLinearArrays}, \cite{LoOptimization}, \cite{Shamonina_superdirectivity_2015}
\begin{equation}\label{eq:Dmatrix}
	\begin{split}
		D(\theta,\phi)
		&=4\pi\frac{\M{I}^{\herm}
		\begin{bmatrix}
			u_{11}(\theta,\phi) & \cdots & u_{1N}(\theta,\phi) \\
			\vdots & \ddots & \vdots \\
	    	u_{N1}(\theta,\phi) & \cdots & u_{NN}(\theta,\phi) \\
		\end{bmatrix}
		\M{I}}
		{\M{I}^{\herm}
		\begin{bmatrix}
			p_{11} & \cdots & p_{1N} \\
			\vdots & \ddots & \vdots \\
			p_{N1} & \cdots & p_{NN} \\
		\end{bmatrix}
		\M{I}}\\
		&=4\pi\frac{\M{I}^{\herm}\M{u}(\theta,\phi)\M{I}}{\M{I}^{\herm}\M{p}\M{I}}
	\end{split}
\end{equation}
where $\herm$ stands for Hermitian transpose, $\M{I}=[I_1 \cdots I_N]^\mathrm{T}$ is a vector of excitation currents and $\M{u}$ and $\M{p}$ are matrices of the normalized mutual radiation intensities $u_{mn}$ and powers $p_{mn}$ respectively. For some simple current distributions $\M{J}$, integrals in  (\ref{eq:Umn}) and (\ref{eq:Pmn}) may be evaluated analytically as is shown later.


\section{End-Fire Array of Two Elementary Dipoles}

Directivity $D$, according to (\ref{eq:Dmatrix}), is further calculated for an array of two elementary dipoles with end-fire radiation. This necessitates the finding of entries $u_{mn}$ and $p_{mn}$ of matrices $\M{u}$ and $\M{p}$.

\subsection{Elementary Dipole}

\begin{figure}[!h]
	\centerline{\includegraphics[width=\columnwidth]{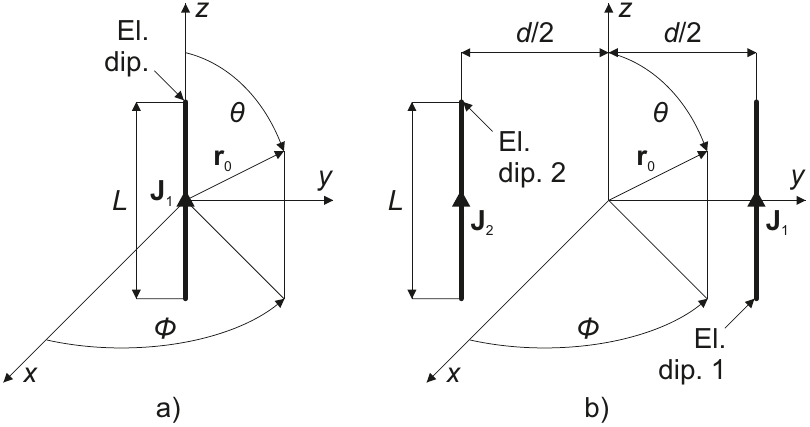}}
	\caption{Geometry: a) elementary dipole, b) array of two elementary dipoles.}
\label{fig0}
\end{figure}

Firstly, let us consider an elementary dipole of length $L\rightarrow0$ in the origin of the coordinates oriented in the $z$-axis with constant current density $\M{J}_1=I_1\delta(x)\delta(y)\M{z}_0=I_1j_{1z}\M{z}_0$, see Fig.~\ref{fig0}~a).

In this case, for intensity $u_{11}$, (\ref{eq:Umn}) becomes
\begin{equation}\label{eq:u11}
	\begin{split}
		u_{11}(\theta,\phi)
		&=\frac{15k^2}{4\pi}\!\!\int\limits_{-L/2}^{L/2}\int\limits_{-L/2}^{L/2}\sin^2\theta\E^{\j k(z-z')\cos\theta}\D z\D z'\\
		&\approx\frac{15k^2}{4\pi}\!\!\int\limits_{-L/2}^{L/2}\int\limits_{-L/2}^{L/2}\sin^2\theta\D z\D z'\\
		&=\frac{15k^2L^2}{4\pi}\sin^2\theta.
	\end{split}
\end{equation}
In the above equation, the Dirac $\delta$-functions reduce 3D volume integrals from (\ref{eq:Umn}) to the 1D line integrals and effectively simplify (\ref{eq:delta}), (\ref{eq:Lambda}) to
\begin{equation}\label{}
\Delta-\Delta'=(z-z')\cos\theta,
\end{equation}
\begin{equation}\label{}
	\begin{split}
		\Lambda(\M{r},\M{r}')
		&=\Lambda_{1z,1z}(\M{r},\M{r}')={j}_{1z}(\M{r}){j}_{1z}^*(\M{r}')\sin^2\theta\\
		&=\sin^2\theta
	\end{split}
\end{equation}
and, finally, approximation $z-z'\approx0$ is used for $L\rightarrow0$.

Then, (\ref{eq:Pmn}) for power $p_{11}$  using (\ref{eq:u11}) reads
\begin{equation}\label{eq:p11}
	p_{11}=\frac{15k^2L^2}{2}\int\limits_{0}^{\pi}\sin^3\theta\D\theta=10k^2L^2.
\end{equation}

For the elementary dipole, considering the excitation current $\M{I}=[I_1]$ and using the above found entries of matrices $\M{u}$ and $\M{p}$, directivity $D$ (\ref{eq:Dmatrix}) becomes
\begin{equation}\label{eq:Delem}
	D(\theta)=\frac{15}{10}\sin^2\theta.
\end{equation}
The value of current $I_1$ is insignificant when calculating directivity $D$ since it is ultimately canceled in (\ref{eq:Dmatrix}).

\subsection{Array of Two Elementary Dipoles}

Now, consider an array of two elementary dipoles of length $L\rightarrow0$ oriented in the $z$-axis and spaced in the $x$-coordinate by a distance $d$ with constant current densities $\M{J}_1=I_1\delta(x-d/2)\delta(y)\M{z}_0$ and $\M{J}_2=I_2\delta(x+d/2)\delta(y)\M{z}_0$, see Fig.~\ref{fig0}~b). It is well known that this arrangement produces end-fire radiation if the dipoles are closely-spaced ($d<\lambda/2$) and excited by out-of-phase currents, i.e., $I_1=-I_2=I$ \cite{Kraus_Antennas}.

In this case, the self-intensities $u_{11}$ and $u_{22}$ are equal to (\ref{eq:u11}), i.e., $u_{11}=u_{22}$, since they cannot depend either on the placement in the coordinates, nor on the mutual placement of the dipoles, and due to the dipoles being identical. From (\ref{eq:Umn}) and (\ref{eq:Lambda}), it follows for mutual intensities $u_{12}$ and $u_{21}$ that $u_{12}=u_{21}^*$ and
\begin{equation}\label{eq:u12}
	\begin{split}
		u_{12}(\theta,\phi,s)
		&=\frac{15k^2}{4\pi}\!\!\int\limits_{-L/2}^{L/2}\int\limits_{-L/2}^{L/2}\sin^2\theta\E^{\j k(\Delta-\Delta')}\D z\D z'\\
		&\approx\frac{15k^2}{4\pi}\!\!\int\limits_{-L/2}^{L/2}\int\limits_{-L/2}^{L/2}\sin^2\theta\E^{\j kd\sin\theta\cos\phi}\D z\D z'\\
		&=\frac{15k^2L^2}{4\pi}\sin^2\theta\E^{\j kd\sin\theta\cos\phi}\\
		&=u_{11}(\theta,\phi)\E^{\j s\sin\theta\cos\phi}.
	\end{split}
\end{equation}
In the above equation, the Dirac $\delta$-functions reduce 3D volume integrals from (\ref{eq:Umn}) to the 1D line integrals and effectively simplify (\ref{eq:delta}), (\ref{eq:Lambda}) to
\begin{equation}\label{}
	\Delta-\Delta'=d\sin\theta\cos\phi+(z-z')\cos\theta,
\end{equation}
\begin{equation}\label{}
	\begin{split}
		\Lambda(\M{r},\M{r}')
		&=\Lambda_{1z,2z}(\M{r},\M{r}')={j}_{1z}(\M{r}){j}_{2z}^*(\M{r}')\sin^2\theta\\
		&=\sin^2\theta
	\end{split}
\end{equation}
and, finally, the approximation $z-z'\approx0$ for $L\rightarrow0$ and a comparison with (\ref{eq:u11}) are used with normalized spacing $s=kd$ being defined.

Powers $p_{11}$ and $p_{22}$ are equal to (\ref{eq:p11}), i.e., $p_{11}=p_{22}$, since intensities $u_{11}$ and $u_{22}$ are equal to (\ref{eq:u11}). From (\ref{eq:Pmn}), it follows for powers $p_{12}$ and $p_{21}$ that $p_{12}=p_{21}^*$ since it holds true that $u_{12}=u_{21}^*$, for intensities $u_{12}$ and $u_{21}$.
Additionally, the power $p_{12}$ is real, thus, $p_{12}=p_{21}$, see Appendix B. Further, (\ref{eq:Pmn}) for a power $p_{12}$  using (\ref{eq:u12}) reads
\begin{equation}\label{eq:p12}
	p_{12}(s)=\frac{15k^2L^2}{4\pi}\!\!\int\limits_{0}^{2\pi}\int\limits_{0}^{\pi}\sin^3\theta\E^{\j s\sin\theta\cos\phi}\D\theta\D\phi.
\end{equation}
The above integral was evaluated elsewhere \cite{LoOptimization}, \cite{Margetis_superdirectivity_1998}. It is noted that the same result can be obtained by the EMF method \cite{PolivkaNotTooShort}, where all terms containing $z-z'$ are discarded. It yields
\begin{equation}\label{eq:p12res}
	p_{12}(s)=15k^2L^2\left(\frac{\sin{s}}{s}+\frac{\cos{s}}{s^2}-\frac{\sin{s}}{s^3}\right).
\end{equation}

For the given array, considering the out-of-phase excitation currents $\M{I}=[I,-I]^\T{T}$ with magnitude $I$ and using the above found entries of matrices $\M{u}$ and $\M{p}$, directivity $D$ (\ref{eq:Dmatrix}) becomes
\begin{equation}\label{eq:D+1-1general}
	D(\theta,\phi,s)=\frac{3\sin^2\theta(1-\cos{(s\sin\theta\cos\phi}))}{2-3\left(\frac{\sin{s}}{s}+\frac{\cos{s}}{s^2}-\frac{\sin{s}}{s^3}\right)}.
\end{equation}
The value of current $I$ is insignificant when calculating directivity $D$ since it is ultimately canceled in (\ref{eq:Dmatrix}). Further, considering the spacing $d<\lambda/2$,  the maximal (end-fire) radiation occurs for the direction $(\theta=90\degree,\phi=0\degree)$ and the corresponding directivity $D$ is
\begin{equation}\label{eq:D+1-1max}
	D(90\degree,0\degree,s)=\frac{3(1-\cos{s})}{2-3\left(\frac{\sin{s}}{s}+\frac{\cos{s}}{s^2}-\frac{\sin{s}}{s^3}\right)}
\end{equation}
with the limit $15/4=3.75$ ($5.74$ dBi) for the spacing $d\rightarrow0$ (i.e., $s\rightarrow0$).

The denominator of (\ref{eq:D+1-1general}) and (\ref{eq:D+1-1max}) represents the interaction of the self- and mutual powers $p_{11}$ and $p_{12}$ and is the leading function describing the behavior of quality factor $Q$ of this array \cite{LoOptimization}, \cite{HazdraCapekEichlerMazanek_DipoleRadiationAboveGroundPlane}, which behaves as
\begin{equation}\label{eq:Qkd}
Q(s)\varpropto\frac{1}{p_{11}-p_{12}(s)}.
\end{equation}

\section{Optimal Excitation and Superdirectivity}
In directivity $D$ (\ref{eq:D+1-1general}), the currents $\M{I}=[I,-I]^\T{T}$ are considered. However, the general expression of directivity $D$ (\ref{eq:Dmatrix}) is a quadratic form in terms of the currents $\M{I}$ and can be used to find their optimum $\M{I}_{\T{opt}}$ which maximizes directivity $D$ for a given direction $(\theta,\phi)$ and spacing $s$ by solving the related weighted eigenvalue equation \cite{Harrington_FieldComputationByMoM}
\begin{equation}\label{eq:Deig}
	4\pi\M{u}\M{I}_\mathrm{opt}=D\M{p}\M{I}_\mathrm{opt}.
\end{equation}
Surprisingly, in this particular case, the currents $\M{I}_\mathrm{opt}=\left [I_{1,\T{opt}},I_{2,\T{opt}}\right ]^\T{T}$ can be found analytically by following the procedure in \cite{UzsokySolymar_TheoryOfSuperDirectiveLinearArrays}, \cite{Shamonina_superdirectivity_2015}. They are given by solution
\begin{equation}\label{eq:Iopt}
	\M{I}_\mathrm{opt}(\theta,\phi,s)=\frac{1}{4\pi}\M{p}^{-1}(s)\M{V}(\theta,\phi,s)
\end{equation}
where
\begin{equation}\label{eq:V}
	\M{V}(\theta,\phi,s)=
	\begin{bmatrix}
		\E^{\j s/2\sin\theta\cos\phi}\sin\theta\\
		\E^{-\j s/2\sin\theta\cos\phi}\sin\theta
	\end{bmatrix}
	.
\end{equation}
Thus, the currents $\M{I}_\mathrm{opt}$ (\ref{eq:Iopt}) can be written with the help of the previously found matrix $\M{p}$
 \begin{equation}\label{eq:Ioptresult}
	\M{I}_\mathrm{opt}(\theta,\phi,s)=
		\begin{bmatrix}
		I\E^{\j \alpha(\theta,\phi,s)/2}\\
		I\E^{-\j \alpha(\theta,\phi,s)/2}
	\end{bmatrix}
\end{equation}
where magnitude $I$ is the same for currents $I_{1,\T{opt}}$ and $I_{2,\T{opt}}$ and $\alpha$ is their phase difference, which reads
\begin{equation}\label{eq:Ioptphasegeneral}
	\alpha(\theta,\phi,s)=-s\sin\theta\cos\phi\\
	+2\arg\left(\rho_{\T{Re}}(\theta,\phi,s)+\j\rho_{\T{Im}}(\theta,\phi,s)\right)
\end{equation}
where
\begin{equation}\label{}
	\begin{split}
		\rho_{\T{Re}}(\theta,\phi,s)=&2\cos{(s\sin\theta\cos\phi)}\\
		&-3\left(\frac{\sin{s}}{s}+\frac{\cos{s}}{s^2}-\frac{\sin{s}}{s^3}\right),
	\end{split}
\end{equation}
\begin{equation}\label{}
	\rho_{\T{Im}}(\theta,\phi,s)=2\sin{(s\sin\theta\cos\phi)}.
\end{equation}

\begin{figure}[!h]
	\centerline{\includegraphics[width=85mm]{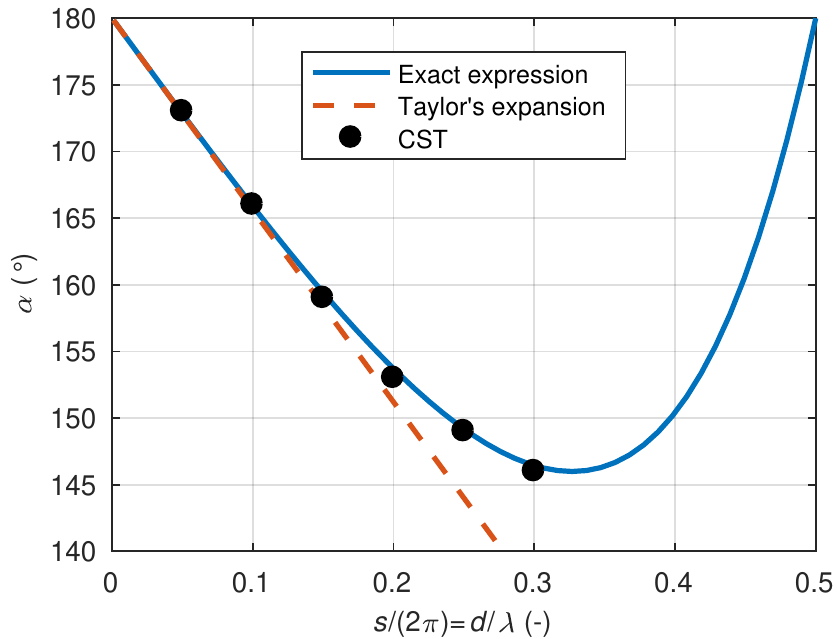}}
	\caption{Phase difference of optimal excitation currents for maximal directivity of the end-fire radiation of an array of two elementary dipoles: exact expression (blue-solid), Taylor's expansion (red-dashed), CST simulation (black-dot).}
	\label{fig1}
\end{figure}

For the given array, considering the optimal excitation currents $\M{I}_{\T{opt}}$ (\ref{eq:Ioptresult}) and using the above found entries of matrices $\M{u}$ and $\M{p}$, directivity $D$ (\ref{eq:Dmatrix}) becomes
\begin{equation}\label{}
	D(\theta,\phi,s)=\frac{3\sin^2\theta(\cos{\alpha}+\cos{(s\sin\theta\cos\phi)})}{2\cos{\alpha}+3\left(\frac{\sin{s}}{s}+\frac{\cos{s}}{s^2}-\frac{\sin{s}}{s^3}\right)}.
\end{equation}
This relation express the maximum directivity $D$ for the given direction $(\theta,\phi)$ and spacing $s$ which is achieved by the excitation of the given array by the currents $\M{I}_{\T{opt}}$ set for the direction $(\theta,\phi)$ and spacing $s$ according to (\ref{eq:Ioptresult}).
Further, considering spacing $d<\lambda/2$, the maximal (end-fire) radiation occurs for the direction $(\theta=90\degree,\phi=0\degree)$ and the corresponding directivity $D$ is
\begin{equation}\label{eq:Doptmax}
	D(90\degree,0\degree,s)=\frac{3(\cos{\alpha}+\cos{s})}{2\cos{\alpha}+3\left(\frac{\sin{s}}{s}+\frac{\cos{s}}{s^2}-\frac{\sin{s}}{s^3}\right)}
\end{equation}
where the phase difference $\alpha$ (\ref{eq:Ioptphasegeneral}) is now
\begin{equation}\label{eq:Ioptphase1}
	\begin{split}
		\alpha(90\degree,0\degree,s)=&2\pi-s\\
		&+2\arctan\left(\frac{2\tan{s}}{2-3\left(\frac{\tan{s}}{s}+\frac{1}{s^2}-\frac{\tan{s}}{s^3}\right)}\right).
	\end{split}
\end{equation}
Directivity $D$ (\ref{eq:Doptmax}) has a limit of $21/4=5.25$ ($7.20$ dBi) when spacing $d\rightarrow0$ (i.e., $s\rightarrow0$). Compared to the out-of-phase excitation, this represents an increase by the ``superdirective factor" of $21/15=1.4$ ($1.46$ dB). As seen from Fig.~\ref{fig1}, phase difference $\alpha$ is almost linear for a close spacing $s$.
This motivates its Taylor's expansion, which, by taking the first terms, gives a simple relation
\begin{equation}\label{eq:IoptphaseTaylor}
	\alpha(90\degree,0\degree,s)\approx\pi-{\frac{2}{5}s}.
\end{equation}
Phase difference $\alpha$ (\ref{eq:Ioptphase1}) is notably similar to that obtained numerically by Yaghjian and Altshuler \cite{YaghSmallSupergain}, \cite{Altshuler}.

\begin{figure}[!h]
	\centerline{\includegraphics[width=85mm]{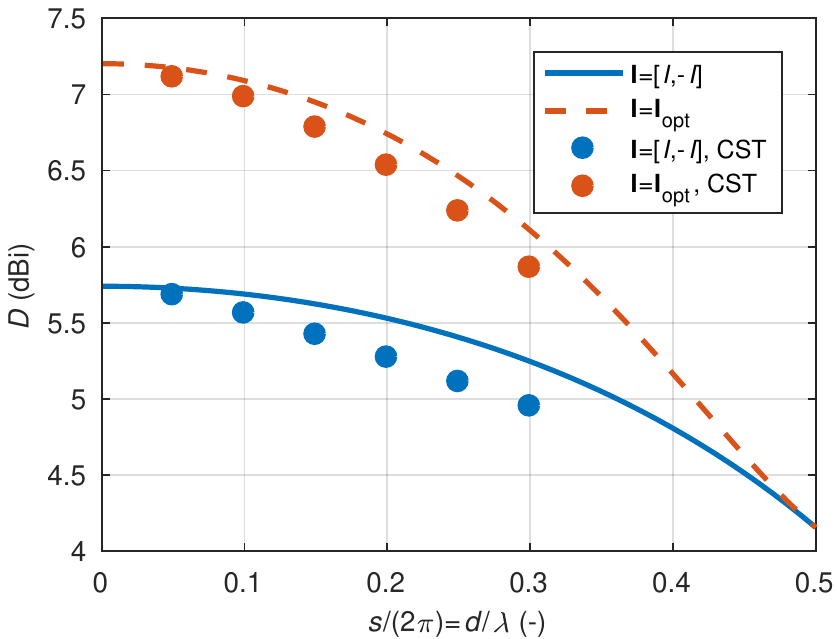}}
	\caption{Directivity of end-fire radiation of an array of two elementary dipoles with out-of-phase (blue-solid) and optimal for maximal directivity (red-dashed) excitation; CST simulation (dot).}
	\label{fig2}
\end{figure}

\begin{figure}[!h]
	\centerline{\includegraphics[width=\columnwidth]{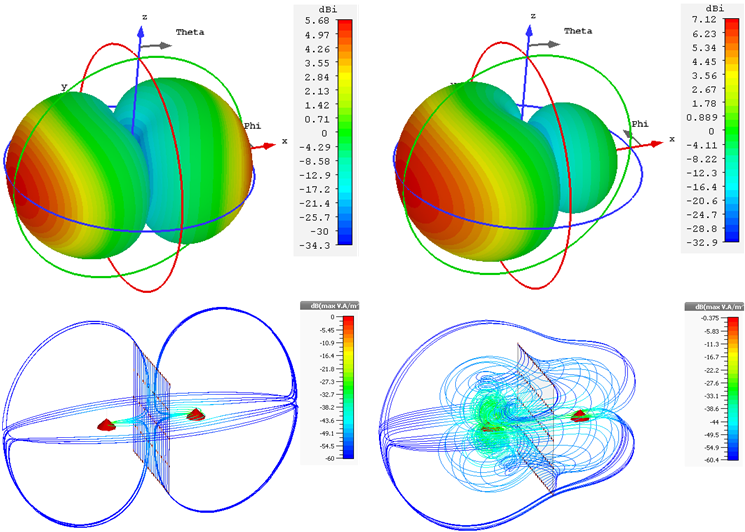}}
	\caption{Radiation pattern for out-of-phase (top-left) and optimal for maximal directivity (top-right) excitation. Streamlines of the Poynting vector are shown below. Spacing $d = 0.1\lambda$.}
	\label{fig3}
\end{figure}

The calculated directivities $D$ (\ref{eq:D+1-1max}) and (\ref{eq:Doptmax}) for both out-of-phase and optimal excitation are shown in Fig. \ref{fig2}.
The results are also validated by the CST simulator \cite{cst}, in which the given array is modeled by two thin dipoles of length $L=\lambda/30$. The optimal phase difference of their excitation currents is, in this case, found manually by varying the phase of the currents in the postprocessing stage and checking the end-fire radiation for maximal directivity. It is seen from Fig.~\ref{fig3} that the radiation patterns of the array for the out-of-phase and optimal excitation are quite distinct. Streamlines of the Poynting vector \cite{ShamoninaStreamlines}, \cite{WarnickStreamlines} are also shown. The interaction between the two dipoles is much stronger for the superdirective case and the power density is more closely bound to the dipoles. Indeed, the fine structure of the power flow is remarkable.

\section{The Uzkov's Limit for Two Isotropic Radiators}
Following proposed approach, Uzkov's  limit $N^2$ for the end-fire directivity of $N$ isotropic radiators \cite{Uzkov} can be verified for $N=2$.

Let us consider an array of two isotropic radiators spaced in the $x$-coordinate by a distance $d$ in the same manner as the elementary dipoles in Fig.~\ref{fig0}~b).

Similarly, as for the case of the array of two elementary dipoles, it holds true for the intensities $u_{11}=u_{22}$, $u_{12}=u_{21}^*$ and
\begin{equation}\label{eq:u12iso}
	u_{12}(\theta,\phi,s)=u_{11}(\theta,\phi)\E^{\j s\sin\theta\cos\phi}.
\end{equation}
However, in comparison to (\ref{eq:u11}), the intensity $u_{11}$ cannot depend on the direction $(\theta,\phi)$ of radiation for the isotropic radiator. From (\ref{eq:Pmn}), the relation of the intensity $u_{11}$ and the power $p_{11}$ can be found
\begin{equation}\label{}
	p_{11}=u_{11}\int\limits_{0}^{2\pi}\int\limits_{0}^{\pi}\sin\theta\D\theta\D\phi=4\pi u_{11},
\end{equation}
\begin{equation}\label{eq:u11iso}
	u_{11}=\frac{p_{11}}{4\pi}.
\end{equation}

It holds true for the powers $p_{11}=p_{22}$ since the intensities $u_{11}$ and $u_{22}$ are equal. Further, (\ref{eq:Pmn}) for the power $p_{12}$  using (\ref{eq:u12iso}) and (\ref{eq:u11iso}) reads
\begin{equation}\label{eq:p12intiso}
	p_{12}(s)=\frac{p_{11}}{4\pi}\!\!\int\limits_{0}^{2\pi}\int\limits_{0}^{\pi}\sin\theta\E^{\j s\sin\theta\cos\phi}\D\theta\D\phi=p_{11}\frac{\sin{s}}{s}.
\end{equation}

For the given array, considering the out-of-phase excitation currents $\M{I}=[I,-I]^\T{T}$ and using the above found entries of the matrices $\M{u}$ and $\M{p}$, directivity $D$ (\ref{eq:Dmatrix}) becomes
\begin{equation}\label{eq:Diso+1-1general}
	D(\theta,\phi,s)=\frac{1-\cos{(s\sin\theta\cos\phi)}}{1-\frac{\sin{s}}{s}}.
\end{equation}
Further, considering the spacing $d<\lambda/2$, the maximal (end-fire) radiation occurs for the direction $(\theta=90\degree,\phi=0\degree)$ and the corresponding directivity $D$ is
\begin{equation}\label{eq:Disomax}
	D(90\degree,0\degree,s)=\frac{1-\cos{s}}{1-\frac{\sin{s}}{s}}
\end{equation}
with limit $3$ ($4.77$ dBi) for the spacing $d\rightarrow0$ (i.e., $s\rightarrow0$).

In this case, the optimal currents $\M{I}_{\T{opt}}$ for the maximal directivity $D$ for a given direction $(\theta,\phi)$ and spacing $s$ can be also found in the manner given by (\ref{eq:Iopt}) and (\ref{eq:V}). They have the same form as (\ref{eq:Ioptresult}) but the phase difference $\alpha$ is now
\begin{equation}\label{eq:IoptphasegeneralISO}
	\alpha(\theta,\phi,s)=-s\sin\theta\cos\phi\\
	+2\arg\left(\rho_{\T{Re}}(\theta,\phi,s)+\j\rho_{\T{Im}}(\theta,\phi,s)\right)
\end{equation}
where
\begin{equation}\label{}
	\rho_{\T{Re}}(\theta,\phi,s)=\cos{(s\sin\theta\cos\phi)}-\frac{\sin{s}}{s},
\end{equation}
\begin{equation}\label{}
	\rho_{\T{Im}}(\theta,\phi,s)=\sin{(s\sin\theta\cos\phi)}.
\end{equation}

For the given array, considering the optimal excitation currents $\M{I}_{\T{opt}}$ (\ref{eq:Ioptresult}) and using the above found entries of the matrices $\M{u}$ and $\M{p}$, directivity $D$ (\ref{eq:Dmatrix}) becomes
\begin{equation}\label{eq:Disooptgeneral}
	D(\theta,\phi,s)=\frac{\cos{\alpha}+\cos{(s\sin\theta\cos\phi)}}{\cos{\alpha}+\frac{\sin{s}}{s}}.
\end{equation}
Further, considering the spacing $d<\lambda/2$, the maximal (end-fire) radiation occurs for the direction $(\theta=90\degree,\phi=0\degree)$ and the corresponding directivity $D$ is
\begin{equation}\label{eq:DoptISO}
	D(90\degree,0\degree,s)=\frac{\cos{\alpha}+\cos{s}}{\cos{\alpha}+\frac{\sin{s}}{s}}
\end{equation}
where the phase difference $\alpha$ (\ref{eq:IoptphasegeneralISO}) is now
\begin{equation}\label{eq:IoptphaseISO}
	\alpha(90\degree,0\degree,s)=-s+2\arctan\left(\frac{\tan{s}}{1-\frac{\tan{s}}{s}}\right)
\end{equation}
with first terms of Taylor's expansion for close spacing $s$
\begin{equation}\label{eq:IoptphaseTaylorISO}
	\alpha(90\degree,0\degree,s)\approx\pi-{\frac{1}{3}s}.
\end{equation}
Directivity $D$ (\ref{eq:DoptISO}) has a limit 4 ($6.02$ dBi) for the spacing $d\rightarrow0$ (i.e., $s\rightarrow0$) corresponding with Uzkov's limit \cite{Uzkov}.

\begin{figure}[!h]
	\centerline{\includegraphics[width=85mm]{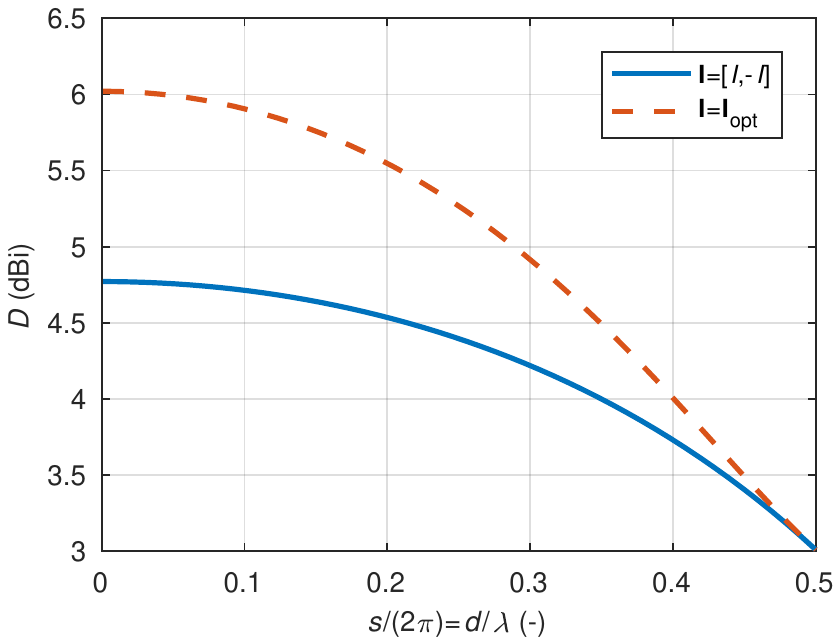}}
	\caption{Directivity of end-fire radiation of an array of two isotropic radiators with out-of-phase (blue-solid) and optimal for maximal directivity (red-dashed) excitation.}
	\label{fig4}
\end{figure}

\begin{figure}[!h]
	\centerline{\includegraphics[width=\columnwidth]{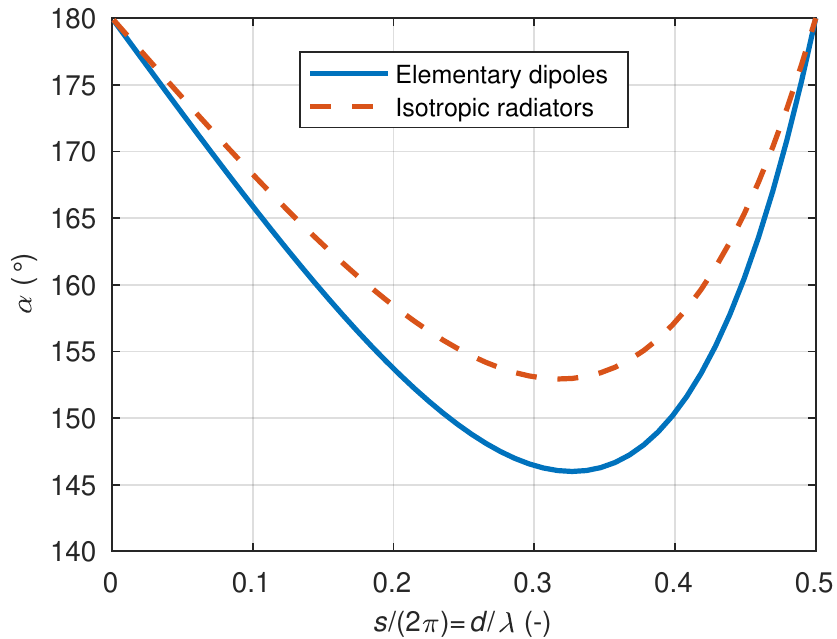}}
	\caption{Comparison of phase difference of optimal excitation currents for maximal directivity of end-fire radiation of an array of two elementary dipoles (blue-solid) and isotropic radiators (red-dashed).}
\label{fig5}
\end{figure}

The calculated directivities $D$ (\ref{eq:Disomax}) and (\ref{eq:DoptISO}) for both out-of-phase and optimal excitation are shown in Fig. \ref{fig4}. A comparison of the phase difference $\alpha$ for the array of the elementary dipoles and isotropic radiators is given in Fig.~\ref{fig5}.

\section{Conclusion}
By using the generalized concept of directivity, an analytical expression for the directivity of an out-of-phase excited array of two closely-spaced elementary dipoles and isotropic radiators was derived. Further, the optimal excitation to maximize the directivity of the arrays for a given direction and spacing was also found analytically with an emphasis on end-fire radiation. Although this is rather an academic case, it illustrates the interesting properties of end-fire radiating arrays. Namely, it shows the dependence of their maximal directivity on the excitation currents, particularly, on their phase difference and the limit of the directivity for the spacing of the array elements approaching zero.
\appendices
\section{Structure of Expression (\ref{eq:Lambda})}
The normalized current density $\M{j}_n$ is usually expressed as a vector $\M{j}_n=[j_{nx},j_{ny},j_{nz}]$ in Cartesian coordinates similarly as the unit vector $\M{r}_0=[\sin(\theta) \cos(\phi), \sin(\theta) \sin(\phi), \cos(\theta)]$, which determines the direction of radiation. This leads to the expression of (\ref{eq:Lambda}) as
\begin{equation}
  	\begin{split}
		\Lambda=&\Lambda_{mx,nx}+\Lambda_{mx,ny}+\Lambda_{mx,nz}\\
		&+\Lambda_{my,nx}+\Lambda_{my,ny}	+\Lambda_{my,nz}\\
		&+\Lambda_{mz,nx}+\Lambda_{mz,ny}	+\Lambda_{mz,nz}
	\end{split}
\end{equation}
where
\begin{IEEEeqnarray}{lCl}
	\Lambda_{mx,nx}&=&j_{mx}j_{nx}^*(\cos^2\theta\cos^2\phi+\sin^2\phi),\\
	\Lambda_{mx,ny}&=&j_{mx}j_{ny}^*(-\sin^2\theta\cos\phi\sin\phi),\\
	\Lambda_{mx,nz}&=&j_{mx}j_{nz}^*(-\cos\theta\sin\theta\cos\phi),\\
	\Lambda_{my,nx}&=&j_{my}j_{nx}^*(-\sin^2\theta\cos\phi\sin\phi),\\
	\Lambda_{my,ny}&=&j_{my}j_{ny}^*(\cos^2\theta\sin^2\phi+\cos^2\phi),\\
	\Lambda_{my,nz}&=&j_{my}j_{nz}^*(-\cos\theta\sin\theta\sin\phi),\\
	\Lambda_{mz,nx}&=&j_{mz}j_{nx}^*(-\cos\theta\sin\theta\cos\phi),\\
	\Lambda_{mz,ny}&=&j_{mz}j_{ny}^*(-\cos\theta\sin\theta\sin\phi),\\
	\Lambda_{mz,nz}&=&j_{mz}j_{nz}^*\sin^2\theta \label{eq:jmzjmz}.
\end{IEEEeqnarray}
\section{Proof that Expression (\ref{eq:p12}) is Real-Valued}
Power $p_{12}$ (\ref{eq:p12}) can be written as
\begin{equation}\label{eq:p12cossin}
	\begin{split}
		p_{12}(s)=&
		\frac{15k^2L^2}{4\pi}
		\!\!\int\limits_{0}^{\pi}
			\sin^3\theta
			\int\limits_{0}^{2\pi}
				\cos( s\sin\theta\cos\phi)
			\D\phi
		\D\theta\\
		&+\j\frac{15k^2L^2}{4\pi}
		\!\!\int\limits_{0}^{\pi}
			\sin^3\theta
			\underbrace
				{\int\limits_{0}^{2\pi}
					\sin( s\sin\theta\cos\phi)
				\D\phi}
			    _{\Phi(\theta,\phi,s)}
			\D\theta.
	\end{split}
\end{equation}
The function $\Phi$ can be further modified with the help of the properties of the $\sin$ and $\cos$ functions
\begin{equation}\label{}
\begin{split}
	\Phi(\theta,\phi,s)					&=\int\limits_{0}^{\pi}
		\sin( s\sin\theta\cos\phi)
	\D\phi
	+\int\limits_{\pi}^{2\pi}
	\sin( s\sin\theta\cos\phi)
	\D\phi\\
	&=\int\limits_{0}^{\pi}
	\sin( s\sin\theta\cos\phi)
	\D\phi
	+\int\limits_{0}^{\pi}
	\sin(-s\sin\theta\cos\phi)
	\D\phi\\
	&=\int\limits_{0}^{\pi}
	\sin( s\sin\theta\cos\phi)
	\D\phi
	-\int\limits_{0}^{\pi}
	\sin(s\sin\theta\cos\phi)
	\D\phi\\
	&=0.
\end{split}
\end{equation}
Thus power $p_{12}$ (\ref{eq:p12cossin}) can be simplified to a formula
\begin{equation}\label{p12cossin}
	p_{12}(s)=
	\frac{15k^2L^2}{4\pi}
	\!\!\int\limits_{0}^{\pi}
		\sin^3\theta
			\int\limits_{0}^{2\pi}
				\cos( s\sin\theta\cos\phi)
			\D\phi
		\D\theta\\
\end{equation}
which produces only real values.

\section*{Acknowledgment}
This work was supported by the Czech Science Foundation under project GA17-00607S Complex Electromagnetic Structures and Nanostructures.

\ifCLASSOPTIONcaptionsoff
  \newpage
\fi



%

\bibliographystyle{IEEEtran}
\bibliography{references}

\begin{thebibliography}{10}
\providecommand{\url}[1]{#1}
\csname url@samestyle\endcsname
\providecommand{\newblock}{\relax}
\providecommand{\bibinfo}[2]{#2}
\providecommand{\BIBentrySTDinterwordspacing}{\spaceskip=0pt\relax}
\providecommand{\BIBentryALTinterwordstretchfactor}{4}
\providecommand{\BIBentryALTinterwordspacing}{\spaceskip=\fontdimen2\font plus
\BIBentryALTinterwordstretchfactor\fontdimen3\font minus
  \fontdimen4\font\relax}
\providecommand{\BIBforeignlanguage}[2]{{%
\expandafter\ifx\csname l@#1\endcsname\relax
\typeout{** WARNING: IEEEtran.bst: No hyphenation pattern has been}%
\typeout{** loaded for the language `#1'. Using the pattern for}%
\typeout{** the default language instead.}%
\else
\language=\csname l@#1\endcsname
\fi
#2}}
\providecommand{\BIBdecl}{\relax}
\BIBdecl

\bibitem{Hansen_Woody_Directional_Antenna_Design}
W.~W. Hansen and J.~R. Woodyard, ``A new principle in directional antenna
  design,'' \emph{Proceedings of the Institute of Radio Engineers}, vol.~26,
  no.~3, pp. 333--345, Mar. 1938.

\bibitem{Bloch_superdirectivity_1953}
A.~Bloch, R.~Medhurst, and S.~Pool, ``A new approach to the design of
  superdirective aerial arrays,'' \emph{Proc. IEE}, vol. 100, pp. 303--314,
  1953.

\bibitem{Uzkov}
A.~I. Uzkov, ``An approach to the problem of optimum directive antennae
  design,'' \emph{Comptes Rendus (Doklady) de l’Academie des Sciences de
  l’URSS}, vol.~53, pp. 35--38, 1946.

\bibitem{Harrington_OnTheGainAndBWofDirectAntennas}
R.~F. Harrington, ``On the gain and beamwidth of directional antennas,''
  \emph{IRE Trans. Antennas Propag.}, vol.~6, no.~3, pp. 219--225, July 1958.

\bibitem{YaghSmallSupergain}
A.~D. Yaghjian, T.~H. O'Donnell, E.~E. Altshuler, and S.~R. Best,
  ``Electrically small supergain end-fire arrays,'' \emph{Radio Science},
  vol.~43, no.~3, pp. 1--13, June 2008.

\bibitem{Haskou}
A.~Haskou, A.~Sharaiha, and S.~Collardey, ``Design of small parasitic loaded
  superdirective end-fire antenna arrays,'' \emph{IEEE Trans. Antennas
  Propag.}, vol.~63, no.~12, pp. 5456--5464, Dec. 2015.

\bibitem{Altshuler}
E.~E. Altshuler, T.~H. O'Donnell, A.~D. Yaghjian, and S.~R. Best, ``A monopole
  superdirective array,'' \emph{IEEE Transactions on Antennas and Propagation},
  vol.~53, no.~8, pp. 2653--2661, Aug. 2005.

\bibitem{Noguchi3el}
A.~Noguchi and H.~Arai, ``3-element super-directive endfire array with
  decoupling network,'' in \emph{2014 International Symposium on Antennas and
  Propagation Conference Proceedings}, Dec. 2014, pp. 455--456.

\bibitem{Best_ImpedanceMatched}
S.~R. Best, E.~E. Altshuler, A.~D. Yaghjian, J.~M. McGinthy, and T.~H.
  O'Donnell, ``An impedance-matched 2-element superdirective array,''
  \emph{IEEE Antennas and Wireless Propagation Letters}, vol.~7, pp. 302--305,
  2008.

\bibitem{Clemente_SuperD}
A.~Clemente, M.~Pigeon, L.~Rudant, and C.~Delaveaud, ``Design of a super
  directive four-element compact antenna array using spherical wave
  expansion,'' \emph{IEEE Transactions on Antennas and Propagation}, vol.~63,
  no.~11, pp. 4715--4722, Nov. 2015.

\bibitem{Kraus_Antennas}
J.~D. Kraus, \emph{Antennas}.\hskip 1em plus 0.5em minus 0.4em\relax
  McGraw-Hill, 1988.

\bibitem{RudgeHandbookVol2}
A.~W. Rudge, \emph{The Handbook of Antenna Design, Vol. 2}.\hskip 1em plus
  0.5em minus 0.4em\relax Institution Of Engineering And Technology, 1983.

\bibitem{UzsokySolymar_TheoryOfSuperDirectiveLinearArrays}
M.~Uzsoky and L.~Solym\'{a}r, ``Theory of super-directive linear arrays,''
  \emph{Acta Physica Academiae Scientiarum Hungaricae}, vol.~6, no.~2, pp.
  185--205, Dec. 1956.

\bibitem{cst}
\BIBentryALTinterwordspacing
{CST Computer Simulation Technology}. (2018) {CST} {MWS}. [Online]. Available:
  \url{http://www.cst.com/}
\BIBentrySTDinterwordspacing

\bibitem{Orfanidis_ElectromagneticWaves}
\BIBentryALTinterwordspacing
S.~J. Orfanidis. Electromagnetic waves \& antennas. [Online]. Available:
  \url{www.ece.rutgers.edu/~orfanidi/ewa}
\BIBentrySTDinterwordspacing

\bibitem{JordanBalmainEM}
E.~C. Jordan and K.~G. Balmain, \emph{Electromagnetic Waves and Radiating
  Systems}.\hskip 1em plus 0.5em minus 0.4em\relax Pearson Education, 2nd ed.,
  2015.

\bibitem{LoOptimization}
Y.~T. Lo, S.~W. Lee, and Q.~H. Lee, ``Optimization of directivity and
  signal-to-noise ratio of an arbitrary antenna array,'' \emph{Proceedings of
  the IEEE}, vol.~54, no.~8, pp. 1033--1045, Aug. 1966.

\bibitem{Shamonina_superdirectivity_2015}
E.~Shamonina and L.~Solymar, ``Maximum directivity of arbitrary dipole
  arrays,'' \emph{IET Microw. Antenna P.}, vol.~9, pp. 101--107, 2015.

\bibitem{Margetis_superdirectivity_1998}
D.~Margetis, G.~Fikioris, J.~M. Myers, and T.~T. Wu, ``Highly directive current
  distributions: General theory,'' \emph{Phys. Rev. E}, vol.~58, p. 2531, 1998.

\bibitem{PolivkaNotTooShort}
M.~Polivka and D.~Vrba, ``Input resistance of electrically short
  not-too-closely spaced multielement monopoles with uniform current
  distribution,'' \emph{IEEE Antennas and Wireless Propagation Letters},
  vol.~11, pp. 1576--1579, 2012.

\bibitem{HazdraCapekEichlerMazanek_DipoleRadiationAboveGroundPlane}
P.~Hazdra, M.~Capek, J.~Eichler, and M.~Mazanek, ``The radiation {Q}-factor of
  a horizontal {$\lambda/2$} dipole above ground plane,'' \emph{IEEE Antennas
  Wireless Propag. Lett.}, vol.~13, pp. 1073--1075, 2014.

\bibitem{Harrington_FieldComputationByMoM}
R.~F. Harrington, \emph{Field Computation by Moment Methods}.\hskip 1em plus
  0.5em minus 0.4em\relax Wiley -- IEEE Press, 1993.

\bibitem{ShamoninaStreamlines}
E.~Shamonina, V.~A. Kalinin, K.~H. Ringhofer, and L.~Solymar, ``Short dipole as
  a receiver: effective aperture shapes and streamlines of the poynting
  vector,'' \emph{IEE Proceedings - Microwaves, Antennas and Propagation}, vol.
  149, no.~3, pp. 153--159, Jun 2002.

\bibitem{WarnickStreamlines}
J.~Diao and K.~F. Warnick, ``Poynting streamlines, effective area shape, and
  the design of superdirective antennas,'' \emph{IEEE Transactions on Antennas
  and Propagation}, vol.~65, no.~2, pp. 861--866, Feb 2017.

\end{thebibliography}


%

\begin{IEEEbiography}[{\includegraphics[width=1in,height=1.25in,clip,keepaspectratio]{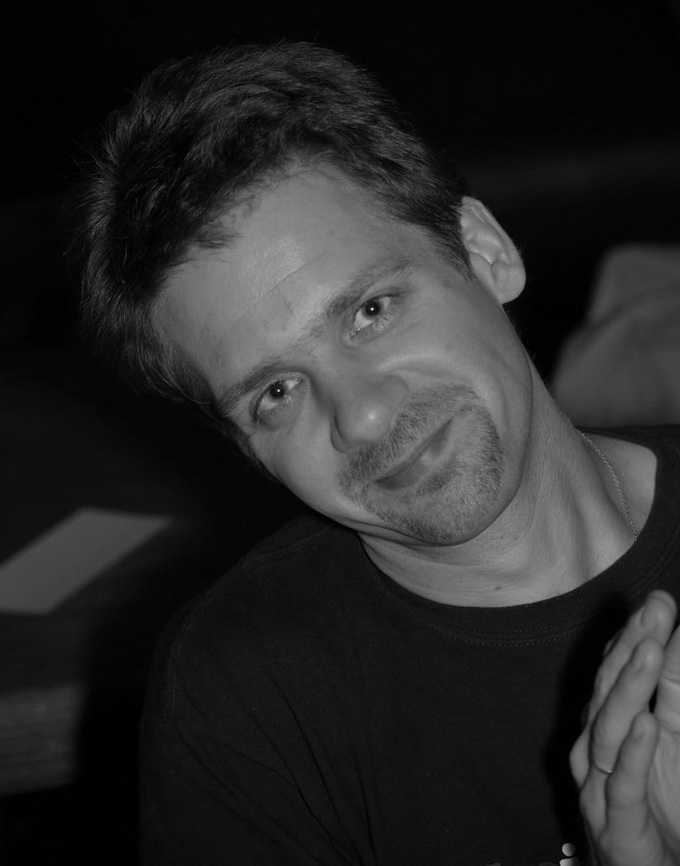}}]{Pavel Hazdra}
was born in Prague, Czech Republic in 1977. He received his M.Sc. and Ph.D. degree in electrical engineering from the Czech Technical University in Prague, Czech Republic, in 2003 and 2009, respectively. Since 2012 he has been associate professor with the Department of Electromagnetic Field at the CTU in Prague. He authored or co-authored more than 25 journal and 30 conference papers. His research interests are in the area of EM/antenna theory, electrically small antennas, reflector antennas and their feeds and antennas for radioamateur purposes.
\end{IEEEbiography}

\begin{IEEEbiography}[{\includegraphics[width=1in,height=1.25in,clip,keepaspectratio]{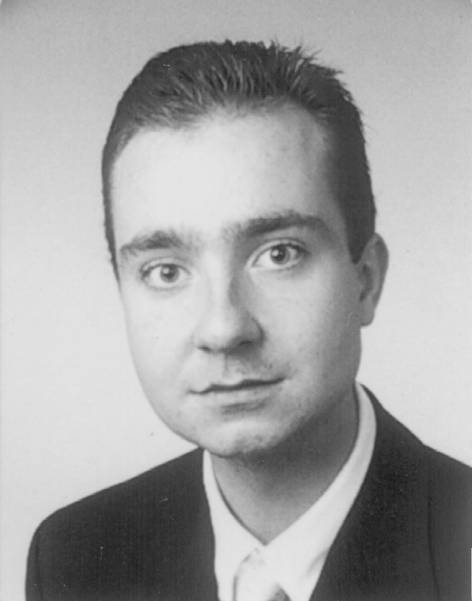}}]{Jan Kracek}
is researcher at the Department of Electromagnetic Field of the Faculty of Electrical Engineering of the Czech Technical University in Prague, Czech Republic. His research interests are theory of electromagnetic field, wireless power transmission and antennas.
\end{IEEEbiography}

\begin{IEEEbiography}[{\includegraphics[width=1in,height=1.25in,clip,keepaspectratio]{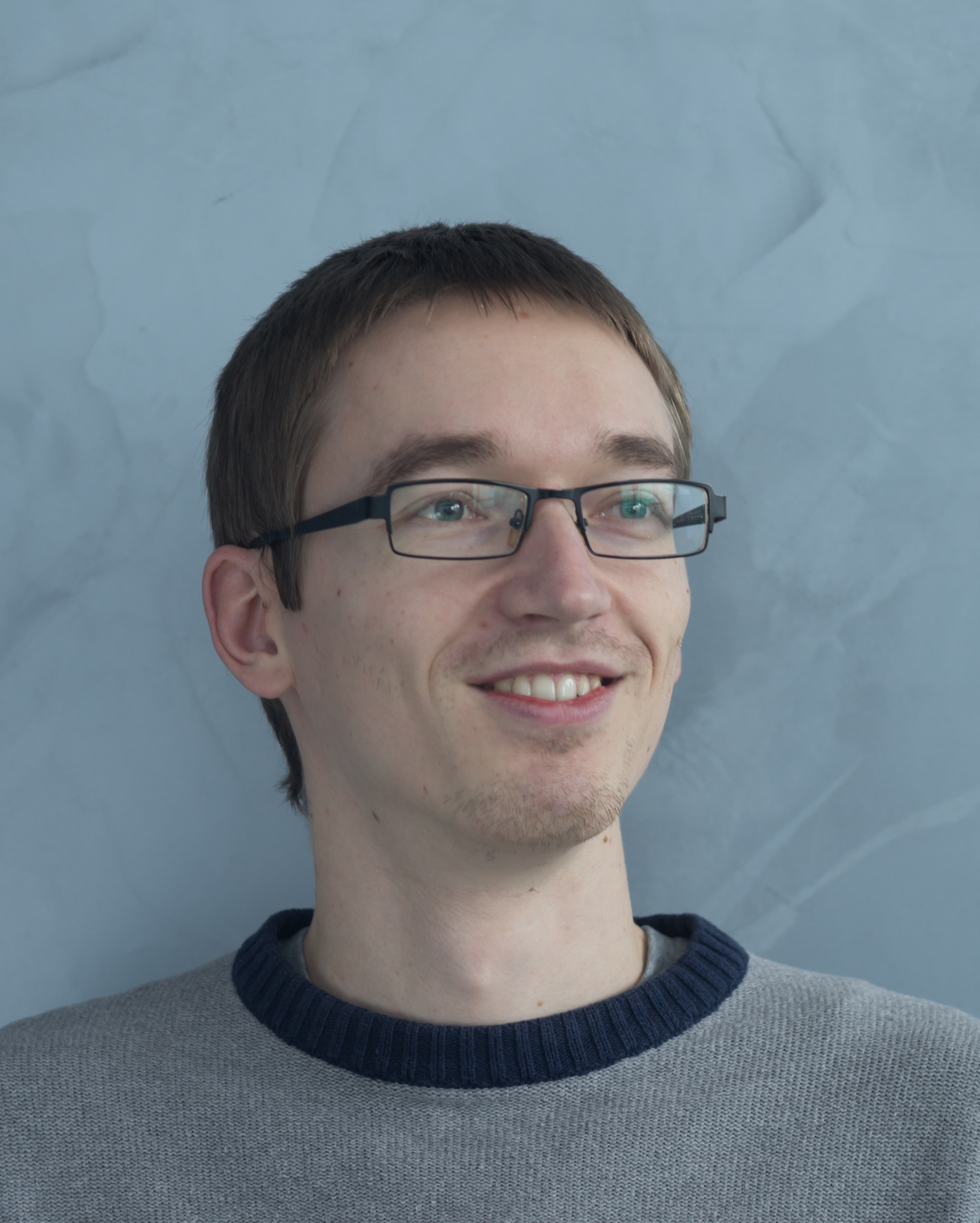}}]{Tomas Lonsky}
received his M. Sc. from Czech Technical University in Prague in 2015. He is working towards his PhD focused on closely-spaced antenna arrays with use of the Theory of Characteristic Modes. In addition he is developing antenna array for 5G mobile networks with a direct connection to optical networks.
\end{IEEEbiography}




\end{document}